\documentstyle[preprint,aps]{revtex}
\begin{document}
\title{Comment on \lq\lq Solution of the Relativistic 
Dirac-Morse Problem \rq\rq}
\author{Arvind Narayan Vaidya\\
Instituto de F\'\i sica - Universidade Federal do Rio de Janeiro \\
Caixa Postal 68528 - CEP 21945-970, Rio de
Janeiro, Brazil
\\
Rafael de Lima Rodrigues\thanks{Permanent address: Departamento de
Ci\^encias Exatas e da Natureza, Universidade Federal da Para\'\i ba,
Cajazeiras - PB, 58.900-000 - Brazil. E-mail to RLR is rafaelr@cbpf.br or
rafael@cfp.ufpb.br and e-mail to ANV is
vaidya@if.ufrj.br.}  
\\ Centro Brasileiro de Pesquisas F\'\i sicas (CBPF)\\
Rua Dr. Xavier Sigaud, 150,CEP 22290-180, Rio de Janeiro, RJ, Brazil}

\maketitle

\newpage

 In a recent letter Alhaidari claims to have formulated and solved the
Dirac-Morse problem \cite{Alha01}. He starts with the Hamiltonian for a 
Dirac particle in an electromagnetic potential and aims to give a 
unified treatment of three problems: Coulomb, oscillator and Morse differing 
only in the choice of gauge. If this were possible gauge invariance would 
imply that the physical content of the theory is not altered by gauge changes.
The known results for the energy spectrum for the first two problems as 
quoted by him contradict this expectation.

In any case if we consider a Hamiltonian that appears in the paper
in a slightly different notation 
\begin{equation}
H= {\mbox{\boldmath $\alpha$}}\cdot({\bf p}-{\bf {\hat  r}}W(r)) +\beta M+V(r)
\end{equation}
where ${\bf {\hat r}} = {{\bf r}\over r}$ and separate variables 
following \cite {BD}
then defining $\Phi=\pmatrix{G_{\ell j}(r)\cr F_{\ell j}(r)\cr}$
we get the radial equation

\begin{equation}
(-i\rho_2{d\over dr} +\rho_1{\kappa\over r}-\rho_2 W-E+V)+M\rho_3)\Phi=0
\end{equation}
where $\rho_i$ are the Pauli matrices and $\kappa=\pm (j+{1\over 2})$ 
for $\ell=j\pm{1\over 2}$,  
which contradicts Alhaidari's equation (1).

Next, Alhaidari's Hamiltonian does not even include that of the 
relativistic oscillator as a special case \cite{Mosh89}. 
The Hamiltonian that satisfies this condition is
\begin{equation}
H= {\mbox{\boldmath $\alpha$}}\cdot({\bf p}-
i\beta{\bf {\hat  r}}W(r)) +\beta M+V(r).
\end{equation}
The resulting radial equation

\begin{equation}
(-i\rho_2{d\over dr} +\rho_1(W+{\kappa\over r})-E+V+M\rho_3)\Phi=0
\end{equation}
does correspond to Alhaidari's equation (1) where the
quantum numbers $\ell$ and $j$ are omitted.
Due to the matrix $\beta$ accompanying $W$ in the Hamiltonian, 
 Alhaidari's interpretation of  the vector $ (V,{\bf {\hat r}} W)$ as an 
electromagnetic potential is incorrect.

 Next, there is no reason for the functions $V(r)$ and $W(r)$
which appear in the Hamiltonian to depend on the angular quantum
numbers which make their appearance only when we separate variables to
solve the Dirac equation. Hence his choice of the constraint (in our notation)

\begin{equation}
W(r)={1\over S}V(r)-{\kappa\over r}
\end{equation}
with both $V$ and $W$ nonzero and $S$ a constant cannot be satisfied. 
Alhaidari could have avoided the mathematical contradiction by taking 
the Hamiltonian to be

\begin{eqnarray}
H= {\mbox{\boldmath $\alpha$}}\cdot({\bf p}-i\beta{\bf {\hat r}}(W(r)+
{K\over r}))
+\beta M+V(r)
\end{eqnarray}
where $ K={\gamma}^0(1+{{\mbox{\boldmath $\Sigma$}}\cdot{\bf L}})$ 
is the Dirac operator,
which leads to the radial equation

\begin{equation}
(-i\rho_2{d\over dr} +W\rho_1-E+V+M\rho_3)\Phi=0
\end{equation}
Applying the transformation $\Phi=e^{-{i\rho_2 \eta}}\hat\Phi $ we get

\begin{equation}
\lbrack-i\rho_2{d\over dr}-(E-V)+\rho_1(W\cos 2\eta-M\sin 2\eta)+\rho_3
(W\sin 2\eta+M\cos 2\eta)\rbrack\hat\Phi=0
\end{equation}
Choosing $ W={V\over \sin 2\eta} $, we get equations (4-5) of 
Alhaidari for $ G_{\ell j}$ and $F_{\ell j}$ leading to  energy levels 
degenerate in $l,j,m$ which is physically uninteresting.   
In the nonrelativistic formulation \cite{Morse29} the radial equation 
for the Morse potential does contain the centrifugal barrier contribution 
for nonzero values of $\ell$.

In conclusion we do not think that the relativistic Morse potential 
problem has been correctly formulated and solved.

\end{document}